# Analysis of Hydrogen Distribution and Migration in Fired Passivating Contacts (FPC)

Mario Lehmann[a,*], Nathalie Valle[b], Jörg Horzel[c], Alisa Pshenova[d], Philippe Wyss[a], Max Döbeli[e], Matthieu Despeisse[c], Santhana Eswara[d], Tom Wirtz[d], Quentin Jeangros[a], Aïcha Hessler-Wyser[a], Franz-Josef Haug[a], Andrea Ingenito[a], Christophe Ballif[a,c]

[a] *Ecole Polytechnique Fédérale de Lausanne (EPFL), Institute of Microengineering (IMT), Photovoltaics and Thin Film Electronics Laboratory, Rue de la Maladière 71b, 2002 Neuchâtel, Switzerland*

[b] *Characterization and Testing Platform (MCTP), Luxembourg Institute of Science and Technology (LIST), Materials Research and Technology Department, 41, rue du Brill, L-4422 Belvaux, Luxembourg*

[c] *CSEM PV-Center, Jaquet-Droz 1, 2002 Neuchâtel, Switzerland*

[d] *Advanced Instrumentation for Ion Nano-Analytics (AINA), Luxembourg Institute of Science and Technology (LIST), Materials Research and Technology Department, 41, rue du Brill, L-4422 Belvaux, Luxembourg*

[e] *ETH Zurich, Laboratory of Ion Beam Physics, Otto-Stern-Weg 5, 8093 Zürich, Switzerland*

\* Corresponding author: mario.lehmann@epfl.ch

**ABSTRACT**

High temperature passivating contacts for c-Si based solar cells are intensively studied because of their potential in boosting solar cell efficiency while being compatible with industrial processes at high temperatures. In this work, the hydrogenation mechanism of fired passivating contacts (FPC) based on c-Si/SiO$_x$/nc-SiC$_x$(p) stacks was investigated. More specifically, the correlation between passivation and local re-distribution of hydrogen resulting from the application of different types of interfacial oxides (SiO$_x$) and post-hydrogenation processes were analyzed. To do so, the applied processing sequence was interrupted at different stages in order to characterize the samples. To assess the hydrogen content, deuterium was introduced (alongside/instead of hydrogen) and secondary ion mass spectroscopy (SIMS) was used for depth profiling. Combining these results with lifetime measurements, the key role played by hydrogen in the passivation of defects at the c-Si/SiO$_x$ interface is discussed. The SIMS profiles show that hydrogen almost completely effuses out of the SiC$_x$(p) during firing, but can be re-introduced by hydrogenation via forming gas anneal (FGA) or by release from a hydrogen containing layer such as SiN$_x$:H. A pile-up of H at the c-Si/SiO$_x$ interface was observed and identified as a key element in the FPC's passivation mechanism. Moreover, the samples hydrogenated with SiN$_x$:H exhibited higher H content compared to those treated by FGA, resulting in higher $iV_{OC}$ values. Further investigations revealed that the doping of the SiC$_x$ layer does not affect the amount of interfacial defects passivated by the hydrogenation process presented in this work. Eventually, an effect of the oxide's nature on passivation quality is evidenced. $iV_{OC}$ values of up to 706 mV and 720 mV were reached with FPC test structures using chemical and UV-O$_3$ tunneling oxides, respectively, and up to 739 mV using a reference passivation sample featuring a ~25 nm thick thermal oxide.

# 1. INTRODUCTION

Over the past decade, the photovoltaic (PV) market has seen a tremendous growth. While the annual installed PV capacity was still below 7 $GW_p$ in 2008, it reached 100 $GW_p$ in 2018 and this trend is expected to continue [1–3]. This evolution was enabled by a continuous increase in solar cell efficiencies [4] and cost reductions for solar cells and PV modules [1]. One of the key factors for high efficiencies is the suppression of recombination losses at the contacts, usually achieved by the deposition of a material that passivates the wafer surface, deactivating defects that act as recombination centers [5,6]. A well-known example is the heterojunction solar cell, where an intrinsic hydrogen-rich amorphous silicon layer is used to passivate interfacial defects. This cell design reaches conversion efficiencies up to 26.7 % in an interdigitated back contacted design, the current world record for single junction c-Si based solar cells [7]. As these heterojunction devices rely on hydrogenated amorphous silicon layers for passivation, they are not compatible with the most common industrial metallization processes that require high-temperatures.

Recently, cells based on so called high temperature passivating contacts (HTPC) have attracted attention thanks to conversion efficiencies >25.5% [8,9] combined with compatibility with the high temperatures (>800 °C) typical of nowadays industrial processes. Most of these passivating contacts are made of a thin (1.2 – 3.6 nm) silicon oxide ($SiO_x$) layer capped with a doped poly-silicon layer (poly-Si). The stack is then annealed at high temperature and subsequently hydrogenated to provide chemical passivation [10–15]. This annealing is usually performed in a tube furnace at temperatures > 800 °C and with heating ramps of 1-10 °C/min, leading to a crystallization of the deposited silicon layers and in-diffusion of dopants into the silicon wafer forming a shallow doped region below the poly-Si/$SiO_x$ stack [16].

In contrast to such approaches, the recently published fired passivating contact (FPC) is fabricated in a single rapid thermal processing (RTP) step, also called firing [17]. This step is used to metallize industrial solar cells [18,19]. Such a process typically requires temperature >750 °C, which are reached with ramps of ~50 °C/s and maintained for a few seconds only. The fabrication of an FPC thus requires a much lower thermal budget than HTPCs based on long annealings. As firing is too short to promote dopant in-diffusion, excellent interface passivation is needed to avoid recombination losses and to achieve high open circuit voltages ($V_{OC}$). Further, the high temperature ramps lead to fast hydrogen effusion, which can lead to blistering. Avoiding such layer delamination is thus a challenge for the FPCs but could be overcome by the addition of carbon into the Si-network [17]. The C content was tuned in order to avert blistering while fostering layer crystallization, which was found to be beneficial for surface passivation and charge carrier extraction. The integration of the FPC as rear hole selective contact, co-fired with a screen printed Ag grid contacting a $POCl_3$ diffused front emitter, resulted in a conversion efficiency of 21.9 % [17].

Hydrogenation is an essential processing step for HTPC (both annealed and fired), during which interfacial defects are passivated, allowing to reach high $V_{OC}$ values. It also plays a key role in surface passivation of many other type of solar cell architectures, and even in bulk quality improvement [20,21]. In this work, the distribution and migration of hydrogen in FPCs is analyzed by means of Secondary Ion Mass Spectrometry (SIMS). The impact of hydrogen re-distribution on surface passivation is studied. A special focus is set on the effect of the various processing steps as well as the influence of the oxide nature on passivation and hydrogen distribution. Deuterium has been incorporated in the samples analyzed by SIMS. The advantage of this being that, in contrast to hydrogen, the deuterium signal is not affected by residual air present in the chamber or humidity adsorbed on the sample surface. Moreover, the detection capability of SIMS is higher for deuterium than for hydrogen.

## 2. EXPERIMENTAL

### 2.1. Fabrication

Symmetrical test structures were fabricated on double side polished (DSP) or shiny etched (SE) p-type float zone (100) wafers. The DSP wafers had a thickness of 280 μm and a resistivity of 3 Ωcm, while the SE wafers, purchased from a different supplier, had a thickness and resistivity of 200 μm and 2 Ωcm, respectively. The first processing step consisted of a wet chemical cleaning, ending with a hot $HNO_3$ treatment (69 %, 80 °C, 10 min) growing a ~1.3 nm thin wet chemical $SiO_x$ layer on the wafer surfaces [22,23]. Next, a ~25 nm thick hydrogenated amorphous a-$SiC_x$(p):H (~2.5 at.% of carbon [17]) layer was deposited by Plasma Enhanced Chemical Vapor Deposition (PECVD) at 200 °C. Subsequently, the samples were fired for 3 s at ~800 °C. During this step, the initially amorphous film (a-$SiC_x$(p):H) crystallizes into nanocrystalline nc-$SiC_x$(p) and hydrogen effuses from that layer. Re-hydrogenation was then performed either via a forming gas anneal (FGA) for 30 min at 500 °C, or via hydrogen diffusion from a ~70 nm thick sacrificial layer of $SiN_x$:H. The latter was deposited at 250 °C in an in-house built PECVD tool and optimized for release of H during a subsequent 30 min hotplate annealing at 450 °C [16]. After this hydrogenation step, the $SiN_x$:H layer was removed in a HF solution. Ellipsometry measurements indicate a refractive index of ~2.0 for these $SiN_x$:H layers. More details about the fabrication process can be found in [17]. Note that the standard hydrogenation route used in this paper is the one by sacrificial $SiN_x$:H layer. FGA was applied only once for comparison.

For the samples to be analyzed by SIMS, deuterium was added into the layers by replacing the $H_2$ gas flows by $D_2$ during the FGA and a-$SiC_x$(p):H/D and $SiN_x$:H/D PECVD processes. Note that during these PECVD depositions, $SiH_4$ and $NH_3$ or trimethylborane (TMB) gas flows were present alongside $D_2$. Thus, both deuterium and hydrogen have been incorporated in these layers. The term hydrogenation is used indifferently whether deuterium is diffused alongside hydrogen for passivation or not.

Due to its higher mass, deuterium has a lower diffusivity than hydrogen [24]. Thus, the kinetics of the hydrogenation process are expected to be different. However, the passivation mechanism should be identical, as the nature of both isotope's bond with Si is the same.

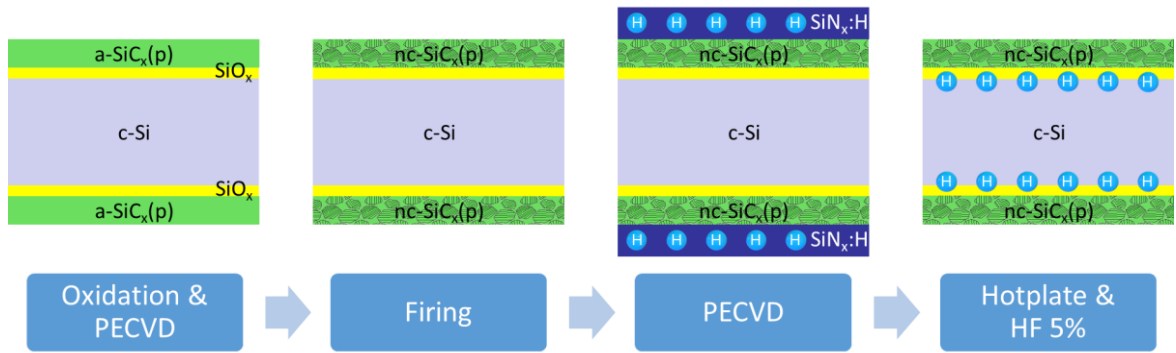

**Fig. 1**: Schematic illustration of the fabrication process using a sacrificial $SiN_x$:H/D layer for hydrogenation. Note that a chemical cleaning was performed prior to the oxidation. Deuterium was incorporated into the layers by replacing the $H_2$ gas flows by $D_2$ in the two mentioned PECVD processes.

Different experiments were performed, investigating the effect of various parameters on the hydrogenation of the FPC. In the first one, the distribution and migration of hydrogen during the processing sequence and its effect on passivation was studied. To do so, deuterium was incorporated into the samples and their passivation and chemical composition were measured at different steps of the processing sequence: 1) after a-$SiC_x$(p) deposition, 2) after firing, and

3) after hydrogenation (FGA or SiN$_x$:H/D). All samples for this study were fabricated on DSP wafers, as a flat surface is needed for SIMS measurements.

In a second experiment, the wet chemical oxide (HNO$_3$) was replaced by a ~25 nm thick thermally grown oxide (90 min in an oxygen ambient at 900 °C, applying N$_2$ ambient while ramping the temperature up and down [25–28]) in order to determine more accurately the location of the deuterium in this layer (1.3 nm being smaller than the SIMS's depth resolution). A reference sample without oxide layer (HF stripping of the SiO$_x$ before the SiC$_x$(p) deposition) was also processed. For both samples SIMS and lifetime measurements were performed in the as-deposited and hydrogenated state, and compared to the previous samples featuring a wet chemical oxide.

In a third experiment, the effect of the interfacial oxide's nature on the passivation quality was investigated. The samples were fabricated on SE wafers, as no SIMS analysis was performed, and their lifetimes measured after hydrogenation (deposition of SiN$_x$:H, hotplate treatment, stripping of the SiN$_x$:H in HF). Three different types of interfacial oxides were compared with each other, namely chemical oxide, grown in hot HNO$_3$ [22,23], UV-Ozone (UV-O$_3$) oxide, grown by exposing the wafer to UV radiation in ambient air (2 min each side) [29–31], and thermal oxide, grown in a tube furnace (grown as detailed above) [25–28]. The thicknesses of the interfacial oxides were ~1.3 nm for the chemical and the UV-O$_3$ oxides and ~25 nm for the thermal oxide, as measured by ellipsometry. Note that the latter sample has to be considered as a reference, as such a stack with a homogenous 25 nm thick SiO$_2$ layer could not be applied as contact. Unfortunately, growing a homogeneous, 1.3 nm thin thermal oxide at 900 °C is very challenging (and reducing the oxidation temperature affects the oxide quality [32]). Thus, a thickness of 25 nm was chosen, in order to enable comparison with the other experiments, where such thick thermal oxides were wanted.

Finally, the effect of the SiC$_x$ layer on the hydrogenation process was studied. To do so, intrinsic, p-type and n-type a-SiC$_x$ layers were deposited on both sides of SE wafers covered with a ~25 nm thick thermal oxide. After firing and hydrogenation, the SiC$_x$ layer was stripped by a selective etch-back in a 20 % KOH solution at 60 °C. Lifetime measurements were performed after each processing step. Note that the samples with an intrinsic SiC$_x$ layer were not fired. Indeed, the SiC$_x$(i) was found to be more prone to blistering than the doped SiC$_x$ layers. Thus, the firing step was replaced by a long hotplate (HP) anneal (7h @ 500 °C) to effuse hydrogen out of the SiC$_x$(i).

**2.2. Characterization**

The passivation quality of the samples was assessed by measuring the photoconductance decay using a Sinton WCT-120 instrument, recording the injection dependent effective minority carrier lifetime ($\tau_{eff}$) and computing the implied open circuit voltage ($iV_{OC}$) at 1 sun [33–35], implementing the Auger correction published by Richter *et al.* [36]. The dark saturation current density ($J_0$) was extracted from this data according to the method published by Kimmerle *et al.* [37]. The $J_0$ values are given per wafer side. From the lifetime, the effective surface recombination velocity ($S_{eff}$) was computed according to Sproul's equation [38]. The value of the diffusivity needed for this computation was determined with the help of PV Lighthouse's mobility calculator [39].

The chemical composition of the layers was measured by SIMS, using a CAMECA SC-Ultra instrument with a 1 keV Cs$^+$ primary ions bombardment. Deuterium was analyzed as D$^-$ and DCs$_2^+$. Ions were collected from an area of 60 μm in diameter, with a depth resolution of ~4 nm (not element dependent) [40]. A selection of samples was further characterized by Rutherford Backscattering Spectrometry (RBS) with a 2 MeV He ion beam [41]. Measurements were performed at the ETH Laboratory of Ion Beam Physics using a silicon PIN diode detector under 168°. The hydrogen and deuterium content of the samples' layers was determined by Elastic

Recoil Detection Analysis (ERDA) under 30° using a 2 MeV He beam and the absorber foil technique [41]. The collected RBS data was analyzed by the RUMP code [42]. Note that the depth resolution of H in Si of this measurement technique is about 50 nm [41]. Our layers being thinner than that, the hydrogen and deuterium contents are given as a surface concentration (at/cm$^2$), corresponding to the total amount of H and D throughout the layer stack.

Layer thicknesses were measured using a UVISEL$^{TM}$ ellipsometer from HORIBA Jobin Yvon S.A.S.

## 3. RESULTS

### 3.1 Hydrogen distribution and migration as a function of the processing step

Looking at the lifetime curves throughout the individual steps of the processing sequence (Fig. 2a), it can be observed that the samples do not reveal any appreciable surface passivation after SiC$_x$ deposition. The firing process then slightly increases their $iV_{OC}$ (< 605 mV, corresponding to $\tau_{eff@10^{15}cm^{-3}}$ < 50 μs). Finally, the post hydrogenation process provides a significant improvement. It is interesting to observe that higher $iV_{OC}$ values were reached when the hydrogenation was done with a sacrificial SiN$_x$:H, rather than via FGA. SiN$_x$:H hydrogenation resulted in $iV_{OC}$ values up to 693 mV ($\tau_{eff@10^{15}cm^{-3}}$ = 950 μs, $J_0$ = 22 ± 5 fA/cm$^2$), whereas FGA treated samples reached only 649 mV ($\tau_{eff@10^{15}cm^{-3}}$ = 190 μs). To gain a deeper understanding of these $iV_{OC}$ trends, deuterium profiles were measured by SIMS (Fig. 2b), analyzing negative secondary ions (high sensitivity to deuterium). First of all, it can be noticed that the deuterium content in the a-SiC$_x$(p):H layer in the as deposited state is high and homogenous. Nevertheless, its $iV_{OC}$ is low due to the defective nature of the SiO$_x$/c-Si interface. During firing, deuterium effuses out of the SiC$_x$(p) and its concentration drops below the detection limit of the SIMS. Finally, the hydrogenation results in an increase of the deuterium content in the SiC$_x$(p) and a strong peak at the position of the SiO$_x$ layer. As expected from lifetime results in Fig. 2a, the hydrogenation by a sacrificial SiN$_x$:H layer introduces more deuterium than the FGA, explaining the observed trends. This is consistent with the work by Lelièvre *et al.* and Dekkers *et al.*, showing that part of the hydrogen released from SiN$_x$:H is in its atomic form, which diffuses more rapidly than molecular hydrogen from FGA [43,44]. Other parameters potentially affecting the hydrogen diffusion are the process temperature (450 °C for the hotplate treatment, vs. 500 °C for the FGA) and the concentration of hydrogen in the source (~18 at.% for SiN$_x$:H, vs. ~4 at.% for FGA).

The total deuterium concentration in the layer, measured by He ERDA, is given in Fig. 2b for the as-deposited, fired and hydrogenated (by a sacrificial SiN$_x$:H/D layer) samples: (2.0 ± 0.4)·10$^{15}$ at/cm$^2$, < 0.1·10$^{15}$ at/cm$^2$ and (0.4 ± 0.2)·10$^{15}$ at/cm$^2$, respectively. Besides the deuterium introduced through a D$_2$ gas flow, there is also hydrogen incorporated into the layers through precursor gases like SiH$_4$ and TMB. The hydrogen content measured by He ERDA for the as-deposited, fired and hydrogenated (by a sacrificial SiN$_x$:H/D layer) samples were (44 ± 5)·10$^{15}$ at/cm$^2$, (3 ± 1)·10$^{15}$ at/cm$^2$ and (5.8 ± 0.6)·10$^{15}$ at/cm$^2$, respectively. Note that these values could be biased by adsorption of humidity on the sample surfaces before ERDA measurements. This effect could explain the fact that the measured hydrogen content is above the detection limit after firing, whereas it drops below this limit for deuterium. Assuming a background signal of 3·10$^{15}$ at/cm$^2$ of hydrogen, a H/D ratio of ~7 is measured after hydrogenation. He ERDA measurements on the SiN$_x$:H layer reveal a H/D ratio of ~1.6, indicating a faster diffusion for hydrogen than for deuterium, in agreement with literature [24]. The as-deposited sample displays a homogeneous hydrogen and deuterium distribution corresponding to a total combined concentration of H + D of (1.8 ± 0.4)·10$^{22}$ at/cm$^3$, i.e. >25 at.% according to [45]. This amount was found to be much lower after firing and hydrogenation, as the layer crystallized, containing thus less structural defects to be hydrogenated.

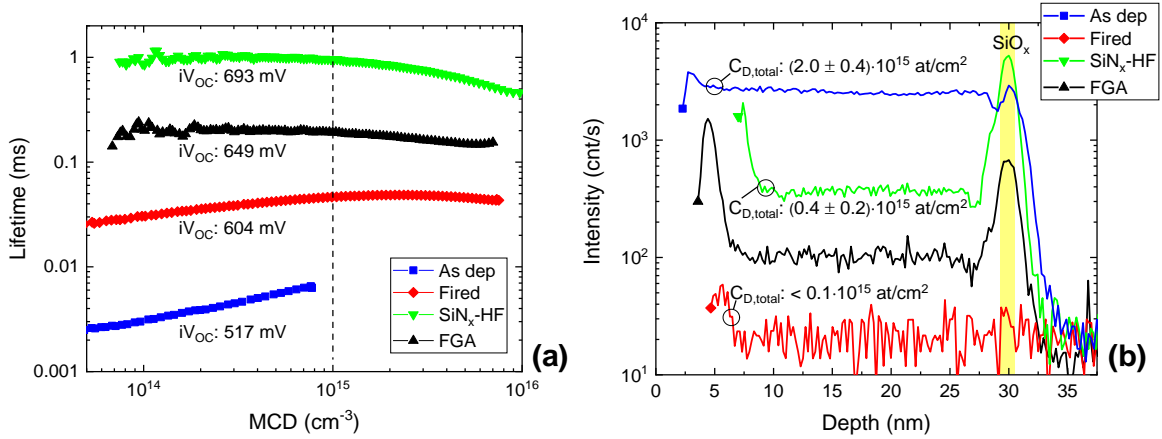

**Fig. 2**: (a): Minority carrier lifetime curves of selected samples at various processing stages, namely as deposited, fired, FGA, SiN$_x$:H hydrogenation (after deposition, hotplate anneal and removal of the SiN$_x$ layer in HF), as a function of minority carrier density (MCD). The dashed line marks the MCD of $1 \cdot 10^{15} \text{cm}^{-3}$ at which $\tau_{eff@10^{15}cm^{-3}}$ in the text are determined. (b): Deuterium depth profiles (D$^-$ SIMS intensities) for samples at different processing stages. The total concentration of deuterium within the samples ($C_{D,total}$), measured by He ERDA, is given for the as deposited, fired and SiN$_x$:H hydrogenated samples.

### 3.2. Hydrogen distribution and migration for different thicknesses of the interfacial oxide

The SIMS depth profiles showed that the deuterium accumulates mainly at the position of the SiO$_x$ layer, which is in agreement with the work performed by Schnabel *et al.* [46] and Dingemans *et al.* [47] using Al$_2$O$_3$:D as deuterium donor layer. However, as the thickness of the chemical oxide is lower than the depth resolution of the SIMS (~4 nm), no conclusion about the exact location of the deuterium can be drawn from these measurements. For a deeper understanding of deuterium accumulation after the hydrogenation process, we analyzed the SIMS profile of a sample grown with a thick thermal oxide, and compared it to those of samples with a thin chemical tunneling oxide and no oxide (Fig. 3). For these measurements, DCs$_2^+$ secondary ions were analyzed, as this mode is less prone to matrix effects and thus a more suited approach to compare signals coming from different materials.

As shown in Fig. 3a, the sample without the interfacial SiO$_x$ has a similar amount of D in the a-SiC$_x$(p) than the other samples, in both as-deposited and hydrogenated states. Despite that, surface passivation is poor ($iV_{OC}$ < 600 mV after hydrogenation). The comparison with Fig. 3b illustrates that it is the previously observed accumulation of D at the SiO$_x$ layer that enables high $iV_{OC}$ values. Finally, Fig. 3c shows that after hydrogenation, D accumulates at both SiO$_2$/nc-SiC$_x$(p) and SiO$_2$/c-Si wafer interfaces whereas its concentration is low within the SiO$_2$. Such results are in agreement with the hypothesis that hydrogen accumulates at defective interfaces to passivate defects. In this specific case the H-accumulation at c-Si/SiO$_x$ enables to reach high $iV_{OC}$ values [48,49]. The especially high $iV_{OC}$ value of 728 mV ($\tau_{eff@10^{15}cm^{-3}}$ = 3350 µs, $J_0$ = 1.4 ± 0.5 fA/cm$^2$) obtained for the sample with the thermal oxide layer indicates potential for improvement for the thin interfacial oxides. It is also interesting to note that the thick thermal oxide layer provides much better passivation in the as deposited state than the chemical oxide.

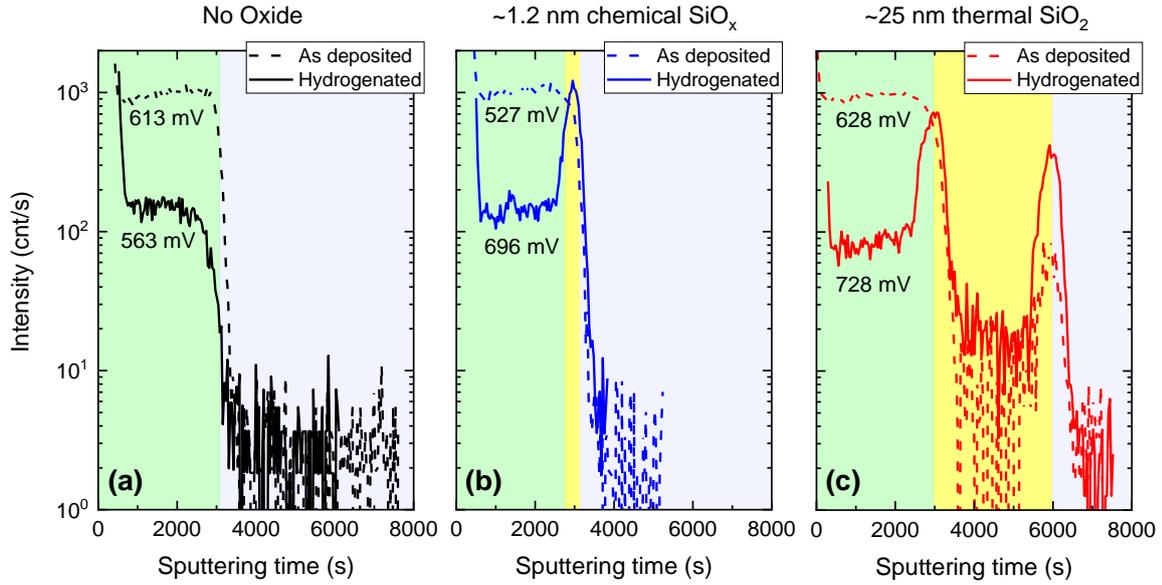

**Fig. 3:** Deuterium profiles ($DCs_2^+$ SIMS intensities) for samples with (a) no interfacial oxide, (b) a chemical tunneling oxide and (c) a thick thermal oxide, in the as deposited state (dashed line) or after hydrogenation (solid line). The yellow area indicates the region of the oxide. The purple region on the right side of the graph corresponds to the c-Si wafer, whereas the a-SiC$_x$(p) layer is located in the green region on the left. The $iV_{OC}$ values of the samples (in mV) are given below the lines.

### 3.3. Effect of the interfacial oxide's nature on the passivation quality

In this section the impact of various interfacial oxides on the passivation is studied. To do so, SiC$_x$(p) layers were deposited on SE wafers covered with a chemical oxide (~1.3 nm), a UV-O$_3$ oxide (~1.3 nm) or a thermal oxide (25 nm). The measured lifetime curves as a function of the injection level are shown in Fig. 4.

A first observation is a trend to higher $iV_{OC}$ values when moving from chemical, to UV-O$_3$, to thermal oxide: 706 mV, 720 mV and 737 mV, respectively, corresponding to $\tau_{eff@10^{15}cm^{-3}}$ of 555 μs, 1100 μs and 3170 μs, and $J_0$ of 8.3 ± 0.6 fA/cm², 7.5 ± 0.3 fA/cm² and 2.3 ± 0.5 fA/cm². This improvement is believed to be linked to changes in the oxide's chemistry, which, according to literature, becomes closer to the stoichiometric ratio of 1:2 when switching from a chemical, to a UV-O$_3$, and then to a thermal oxide [29,50].

Note that while the samples with chemical and UV-O$_3$ oxide can be compared directly, as the type of interfacial oxide was the only parameter varied, care has to be taken when comparing them with the sample with thermal oxide as the latter is thicker and has a different thermal history. Nevertheless, these results show that the nature of this oxide has a major influence on the final passivation quality.

A second observation is that all $iV_{OC}$ values are ~10 mV higher than in the previous experiments, thus exceeding 700 mV even for the samples processed with a chemical oxide (as previously published in [17]). The $J_0$ value decreases from 22 to 8.3 fA/cm² for the samples featuring a chemical oxide on a DSP and SE wafer, respectively, and increases from 1.4 to 2.3 fA/cm² for the DSP and SE samples with a thermal oxide. The reason behind this difference is unclear. Potential factors are the nature of the surface and different bulk lifetimes (as these wafers are provided by different suppliers). Furthermore, it should be mentioned that the computation of $J_0$ becomes inaccurate and dominated by experimental uncertainties when approaching low values (< 4 fA/cm²) [37].

Finally, the 720 mV of iV$_{OC}$ reached for the sample with a UV-O$_3$ tunnelling oxide layer, corresponding to a J$_0$ of 7.5 ± 0.3 fA/cm², confirm the high potential of the FPC.

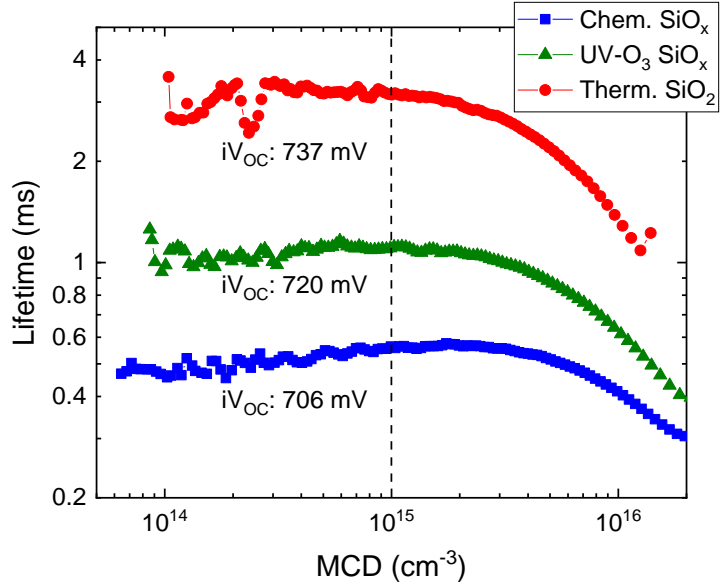

**Fig. 4:** Minority carrier lifetime curves, after hydrogenation, of samples with various interfacial oxides (chemical, UV-O$_3$, thermal) on p-type SE wafers. The dashed line marks the Minority Carrier Density (MCD) of $1 \cdot 10^{15}\,\text{cm}^{-3}$ at which $\tau_{eff@10^{15}cm^{-3}}$ in the text are determined.

### 3.4. Study of the influence of the SiC$_x$ layer's doping on the hydrogenation process

A set of samples with p-type, n-type and intrinsic SiC$_x$ layers deposited by PECVD on thick thermal oxides was prepared to investigate whether the doping of the nc-SiC$_x$ layer affects the hydrogenation process of the FPC, i.e. whether it influences the amount and charge state of the H diffused to the c-Si/SiO$_x$ interface, thus affecting the passivation quality. Such an effect has been reported by Yang *et al.* [51]. The usual processing sequence was completed with a selective etch-back of the SiC$_x$ layer in a KOH solution, in order to eliminate a potential field effect contribution by the doped layer to the surface passivation. Care was taken to selectively etch the partially crystallized SiC$_x$ layer and not the underlying oxide layer, such that the passivation of the interface was not compromised. The results are reported in Fig. 5.

As can be observed, all *iV$_{OC}$* values remain < 690 mV until the SiN$_x$:H deposition. Excellent passivation can then be obtained thanks to the diffusion of hydrogen from the nitride layer towards the c-Si/SiO$_2$ interface on a hotplate, reaching *iV$_{OC}$* values > 740 mV for all SiC$_x$ layers studied here. This value drops slightly (by 4-8 mV) after stripping of the SiN$_x$ layer. Similarly, after etching off the SiC$_x$ layer, only a slight degradation in *iV$_{OC}$* (by 1-8 mV) was observed, the exception being the SiC$_x$(i) which presented local blistering that probably induced inhomogeneous etching and thus locally severe damage of the oxide layer altering the surface passivation, as indicated by an increase in *J$_0$* from 2.3 ± 0.5 fA/cm$^2$ to 13.2 ± 0.9 fA/cm$^2$.

Just after PECVD of the SiN$_x$:H layer, a difference of ~40 mV in *iV$_{OC}$* can be observed between samples featuring a SiC$_x$(p) layer and samples featuring a SiC$_x$(i/n) layer. The origin of this effect is still unclear and requires further investigation.

However, this difference vanishes after hydrogenation, indicating that the final amount of defects passivated by hydrogen (diffused for 30 min at 450 °C) is independent of the layer doping. Further experiments aiming at investigating the impact of the layer doping on the kinetics of this hydrogenation process are required. Moreover, the fact that the SiC$_x$ layer can be removed without major passivation loss indicates that, in the present case of a ~25 nm thick thermal SiO$_2$, the doped SiC$_x$ layer does not contribute to the passivation. Assuming that the fixed charge density in the thermal oxide is low [52,53], the high *iV$_{OC}$* values can be predominantly attributed to the accumulation of hydrogen at the c-Si/SiO$_2$ interface. Whether it is chemical passivation alone or if, and to which extent, interfacial charges play a role, remains an open question.

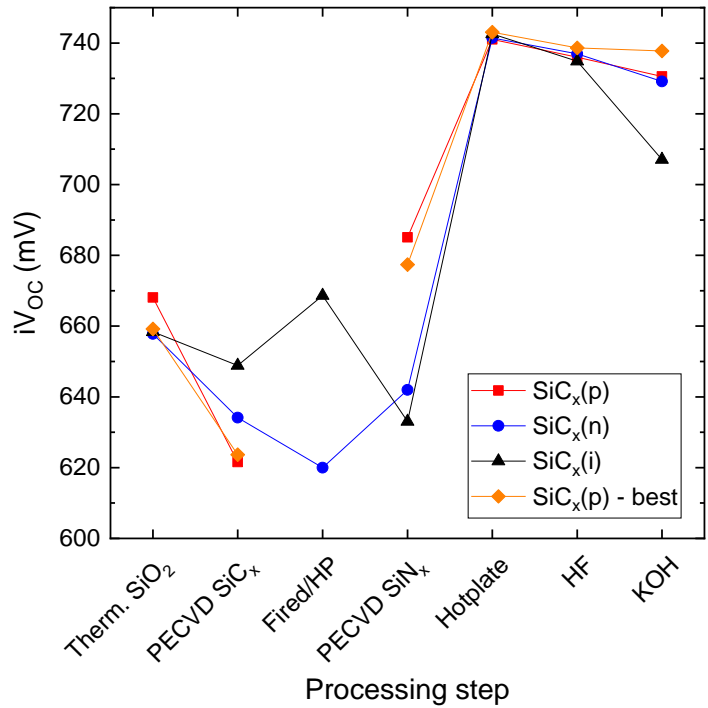

**Fig. 5:** $iV_{OC}$ values as a function of the processing step of samples fabricated with a p-type, n-type or intrinsic $SiC_x$ layer on p-type SE wafers with a 25 nm thick thermal oxide. The values for the best sample processed are also shown. The passivation qualities of the $SiC_x(p)$ samples after firing were too low to measure an $iV_{OC}$.

In samples with ultra-thin tunnelling oxides additional mechanisms might come into play such as superficial changes in carrier concentrations within the wafer, induced by the doped $SiC_x$ layer (leading to band-bending) [17]. But the observations from section 3.1 suggest that hydrogen passivation of the c-Si/$SiO_x$ interface is also the key element to reach high lifetime values with ultra-thin oxides.

In this experiment, $iV_{OC}$ values up to 739 mV after hydrogenation and stripping of the $SiN_x$:H layer were reached, corresponding to a $\tau_{eff@10^{15}cm^{-3}}$ of 3260 µs, a $J_0$ of 2.7 ± 0.7 fA/cm² and a $S_{eff@10^{15}cm^{-3}}$ of 3 cm/s. According to literature, these values correspond to state-of-the-art passivation levels of p-type silicon wafers [36,53–61]. Note also that the present samples have no in-diffused doped region, as stated previously, and that these oxides were grown at 900 °C, a comparably low temperature, and without addition of trichloroethane (TCA). Both, an increased oxidation temperature and an addition of TCA may improve the passivation quality of the thermal oxide [27,62], but also increase the process' complexity.

## 4. CONCLUSION

The combination of lifetime measurements with SIMS analysis elucidated the key role of hydrogen in passivating defects at the c-Si/$SiO_x$ interface to reach high $iV_{OC}$ values. Moreover, it could be observed that hydrogen almost completely effuses out of the $SiC_x(p)$ during firing and is later re-introduced during the hydrogenation step. Performing this hydrogenation step via a $SiN_x$:H sacrificial layer was demonstrated to be more efficient than FGA, which could be correlated with a higher amount of deuterium diffused into the contact and especially to the oxide-wafer interface. Further investigations revealed that in the case of ~25 nm thick thermal oxides, the accumulation of hydrogen at the c-Si/SiO$_2$ interface is the predominant factor enabling excellent passivation levels, and that the doping of the SiC$_x$ layer does not affect the amount of interfacial defects passivated by our hydrogenation process. Furthermore, it was observed that the nature of the interfacial oxide has a major impact on the passivation quality. $iV_{OC}$ values up to 720 mV could be reached using an ultra-thin UV-O$_3$ tunneling oxide, and up to 739 mV on a reference passivation sample using a ~25 nm thick thermal oxide.

**Acknowledgments**

The authors gratefully acknowledge support by the Swiss National Science Foundation (SNF) under Grant no. 200021L_172924/1. The work was co-funded by the Luxembourg National Research Fund (FNR) through grant INTER/SNF/16/11536628 (NACHOS). Xavier Niquille is thanked for the wet chemical cleaning of the wafer and the growing of the UV-O$_3$ oxides. The authors also thank Brahime El Adib (LIST) for his technical assistance with SIMS depth profiling, Christoph Peter (ISC Konstanz) for non-contact corona-Kelvin measurements (not shown), as well as Gizem Nogay and Josua Stückelberger for fruitful discussions.

**Conflict of interest**

The authors declare no conflict of interest.